# Reducing the Gate-Induced Drain-Leakage Current of Carbon-Nanotube Transistors by Subliming Dispersants with Back-End-of-Line Compatible Temperature

*Takumi Inaba[#1], Yuichi Kato[#2*], Yoko Iizumi[2], Yasuhiko Fujita[3], Kazufumi Kobashi[2], Takahiro Morimoto[2], and Toshiya Okazaki[2]*

1 Semiconductor Frontier Research Center, National Institute of Advanced Industrial Science and Technology (AIST), Tsukuba, Ibaraki, 305-8568, Japan

2 Nanocarbon Material Research Institute, National Institute of Advanced Industrial Science and Technology (AIST), Tsukuba, Ibaraki, 305-8565, Japan

3 Research Institute for Sustainable Chemistry, National Institute of Advanced Industrial Science and Technology (AIST), Kagamiyama 3-11-32, Higashihiroshima, Hiroshima, 739-0046, Japan.

**Corresponding Author**

*Y. Kato; National Institute of Advanced Industrial Science and Technology (AIST), Tsukuba, Ibaraki, 305-8565, Japan; Tel: +81-50-3522-7051; E-mail: yuichi.katou@aist.go.jp

## ABSTRACT

Carbon nanotube field-effect transistors (CNTFETs) are promising for monolithic 3D integration with silicon CMOS circuits in back-end-of-line (BEOL) processes. Further reduction in off-state leakage is essential for practical applications. In this study, we develop a low-temperature cleaning process for CNT-deposited wafers using a sublimable CNT dispersant. The process is carried out below 250 °C, making it BEOL-compatible, and does not deteriorate CNT morphology or lattice structure, as confirmed by atomic force microscopy and Raman spectroscopy. As a result, gate-induced drain leakage current originating from residual dispersants is reduced by approximately one order of magnitude. In addition, improvements in on-current, subthreshold slope, and threshold voltage variability are observed after the cleaning process. This study demonstrates a material-enabled approach to improve the transfer characteristics of CNTFETs and provides a practical route toward low-power CNT electronics.

Data transfer between logic circuits and memory in electronic devices dominates the overall device power consumption [1]. The power consumed by data transfer can be mitigated by the monolithic 3D (M3D) integration of multiple logic and memory layers on a single chip. Carbon nanotube field-effect transistors (CNTFETs) are strong candidates for M3D devices because they can be fabricated using back-end-of-line (BEOL)-compatible processes [2-7]. As the primary goal of M3D devices is to reduce power consumption, the CNTFETs employed in these systems must exhibit low power dissipation. The off-state current ($I_{off}$) is a key metric for evaluating transistor-power consumption, particularly when the devices are in standby mode. For comparison, silicon transistors designed for low-power logic circuits are expected to achieve $I_{off}$ values as low as 100 pA/μm [8]. Therefore, research efforts aimed at reducing the $I_{off}$ of CNTFETs below current levels, such as 60 nA/μm [9], are of great importance.

The mechanisms underlying $I_{off}$ in CNTFETs are actively being investigated to achieve its effective reduction [10-14]. From a device physics perspective, $I_{off}$ is generally constrained by the gate-induced drain leakage (GIDL). The major origin of GIDL is band-to-band tunneling (BTBT)[15], which occurs when strong band bending under a high electric field significantly increases the tunneling probability between the valence and conduction bands. To fully understand GIDL mechanisms in practical devices, extrinsic effects should also be considered along with BTBT. Extrinsic effects in CNT research are primarily induced by dispersants. The fabrication of CNTFETs requires the isolation of individual semiconducting CNTs from the CNT aggregates in solution. In this process, the CNT aggregates and dispersants are introduced into a solvent. Poly[9-(1-octyl)-9H-carbazole-2,7-diyl] (PCz)[12,16–18], poly[9,9-di-n-octylfluorenyl-2,7-diyl] (PFO) and its derivatives[19,20] are frequently employed as dispersants for CNTFET fabrication. When the CNT solution is deposited onto wafers, these dispersants remain on the wafers along with CNTs. To mitigate extrinsic effects caused by dispersants, the CNT-deposited wafers are typically cleaned after deposition. Rinsing with an organic solvent is the most common process reported in the literature[12, 16,17, 21-24]. However, owing to wet-processing, both the dispersants and CNTs that adhere to the wafers through van der Waals forces are at risk of being removed during cleaning. This can affect the yield and variability of future heavily scaled CNTFET technologies.

To address these issues, we have developed a novel cleaning process for CNTFETs based on a sublimable dispersant[25]. The sublimable dispersant enables the separation of semiconducting CNTs from CNT aggregates, while allowing the removal of the dispersant from wafers through vacuum annealing at 250 °C. Importantly, the cleaning process is BEOL-compatible and facilitates its integration into M3D device fabrication. Furthermore, it does not involve the wet-processing of CNT-deposited wafers. Therefore, the unintentional peeling of CNTs films, which affects the yield and variability of future heavily scaled CNTFET technologies, can be mitigated. Based on the cleaning process, we demonstrate that the $I_{off}$ value of the CNTFETs is reduced by one order of magnitude towards the development of low-power-consumption M3D devices.

To fabricate CNTFETs, CNTs are generally deposited on a wafer via wet processing[3,5,7,10,12-14, 16-24, 26]. To prepare the CNT solution, a flavin derivative with an octyl side chain (10-octyl-7,8-dimethyl-10H-benzo[g]-pteridine-2,4-dione (FC8)) was employed as a CNT dispersant[25]. Flavin derivatives function as low-molecular-weight nonionic dispersants for single-walled CNTs[27, 28], selectively enrich semiconducting CNTs[29], and can be removed via sublimation[25, 30]. Although a low molecular weight favors sublimation, this feature is inherently in conflict with dispersibility. Achieving a stable dispersion typically requires a sufficiently high molecular weight to allow dispersants to wrap the CNT surfaces effectively[31,32]. Flavin derivatives circumvent this limitation; their supramolecular assemblies, reinforced by intermolecular hydrogen bonding, stabilize CNT dispersions, despite their low molecular weight[28, 33]. FC8 was synthesized as described in the previous study[25]. To prepare the CNT solution, 20 mg of the CNT powder, synthesized using the HipCO method[34], was mixed with 20 mg of FC8 and Toluene (40 ml). The mixture was ultrasonicated and centrifuged.

Figure 1 compares the absorbance spectra of the CNT solution before and after centrifugation as a function of wavelength. After centrifugation, the baseline of the spectra considerably decreased while maintaining light absorption owing to the $E_{11}$ of the semiconducting single-walled CNTs (see supplementary data Fig. S1). The decrease in the baseline indicates the reduction of metallic CNTs and bundled CNTs[35].

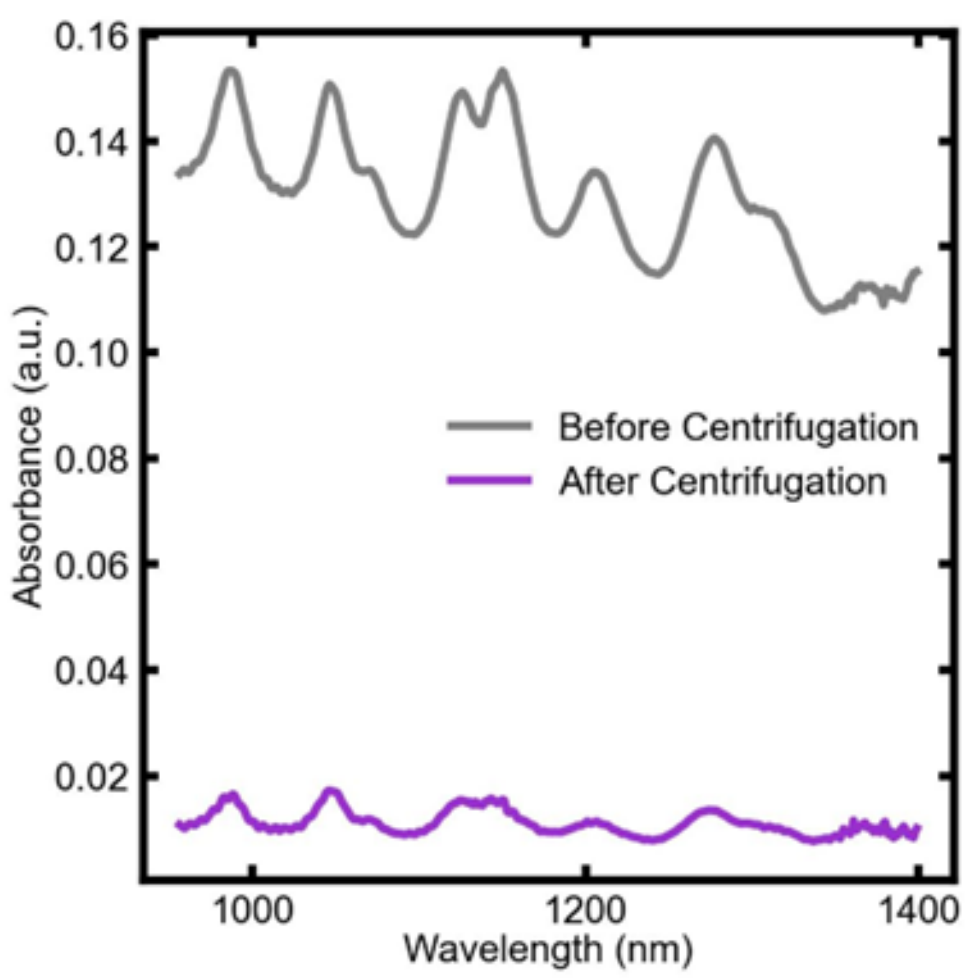

**Fig. 1.** Absorbance spectra of CNT solutions (gray) before and (purple) after centrifugation.

Before fabricating the CNTFETs, we examined the sublimation of FC8 using a BEOL-compatible cleaning process. For this purpose, the CNT solution was spin-coated onto silicon wafers. Atomic force microscopy (AFM) and Raman spectroscopy were performed to examine the CNTs and FC8 on the wafers after spin-coating. Wafers were then annealed under vacuum conditions ($<10^{-3}$ Pa) at 250 °C for 12 h. Finally, AFM and Raman spectroscopy were performed again after cleaning to confirm the FC8 sublimation.

Figure 2 shows a comparison of the AFM images of the wafer with the deposited CNT before and after cleaning. The inset in Fig. 2(a) shows the cross-sectional profile along the white arrow. Before annealing, one-dimensional structures and flakes were observed. As shown in the inset, the height of the thick one-dimensional structure was approximately 8 nm, suggesting that these were likely bundled CNTs. In contrast, the thin one-dimensional structures observed in Fig. 2 were likely isolated CNTs. The flakes in Fig. 2(a) had heights of 1–2 nm and disappeared after cleaning, indicating that these flakes were likely aggregated FC8.

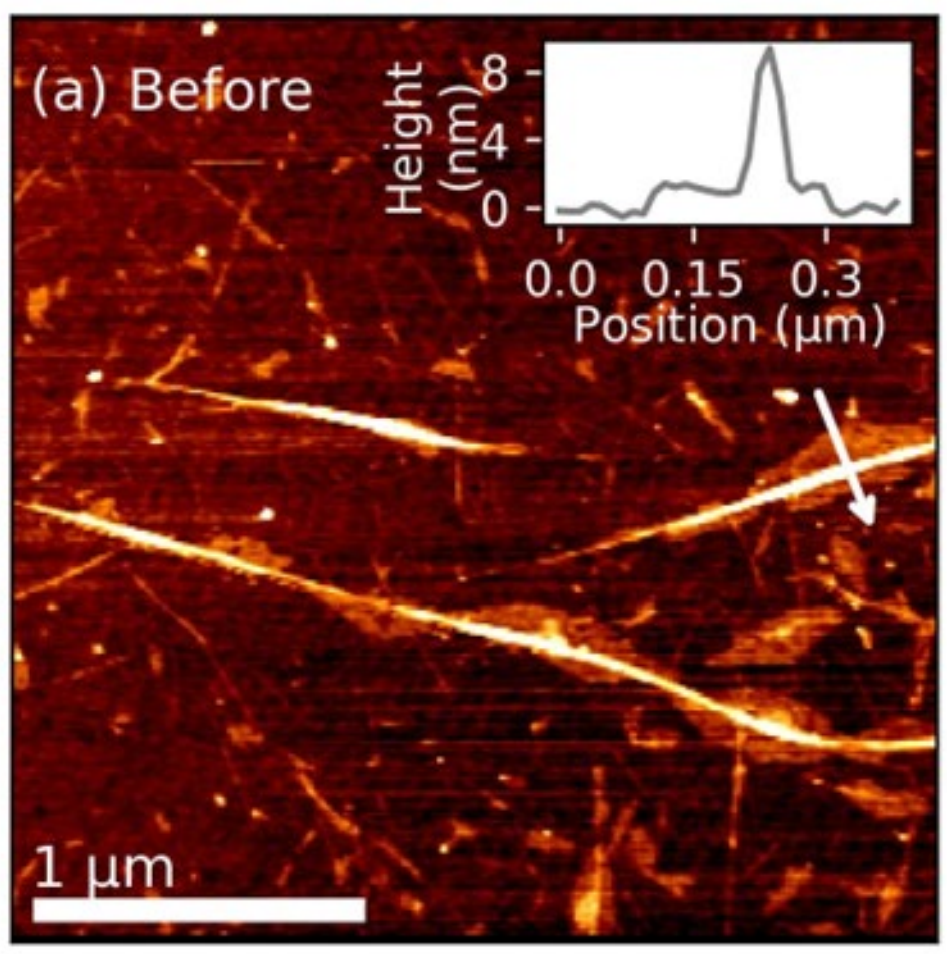

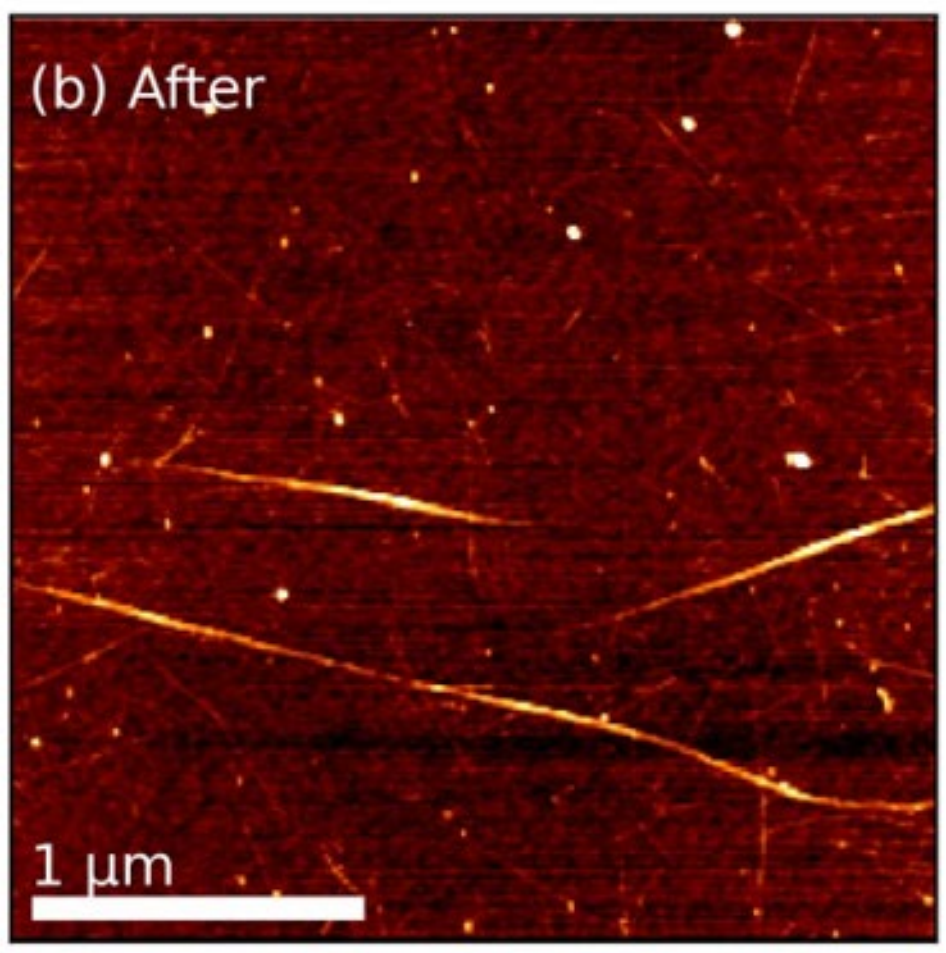

**Fig. 2.** AFM images of a CNT-deposited wafer (a) before and (b) after the annealing. The inset in (a) represents the cross-sectional profile along the arrow in (a).

Based on the results shown in Fig. 2, we conclude that CNTs can be deposited on silicon wafers using the CNT solution, and FC8 can be removed by the BEOL-compatible cleaning process. Furthermore, the positions of the CNTs in Fig. 2(a) and (b) remained unchanged, demonstrating that this cleaning process preserved the CNT-film morphology on the wafers.

Figure 3(a) compares the Raman spectra of the CNT-deposited wafer before (gray) and after (purple) annealing. These two spectra were obtained from the same position on a wafer, which can be confirmed by the identical shape of the radial breathing modes (RBM) at 230 $cm^{-1}$. The peak position of the RBM indicates that the diameter of the CNTs was 1.02 nm[36], which is consistent with the absorbance spectra shown in Fig. 1. Raman mapping (Fig. S2) of the wafer also confirmed that the spectra were obtained from the same position. In Fig. 3(a), the Raman spectrum obtained only from FC8 on the wafer is shown in yellow. Enlarged Raman spectra in the range of 1200–1700 $cm^{-1}$ are shown in Fig. 3(b). Figure 3(b) demonstrates that the FC8-specific Raman peaks around 1500 $cm^{-1}$ were eliminated after the BEOL-compatible cleaning process without increasing the defect-induced Raman mode; i.e. the D-mode.

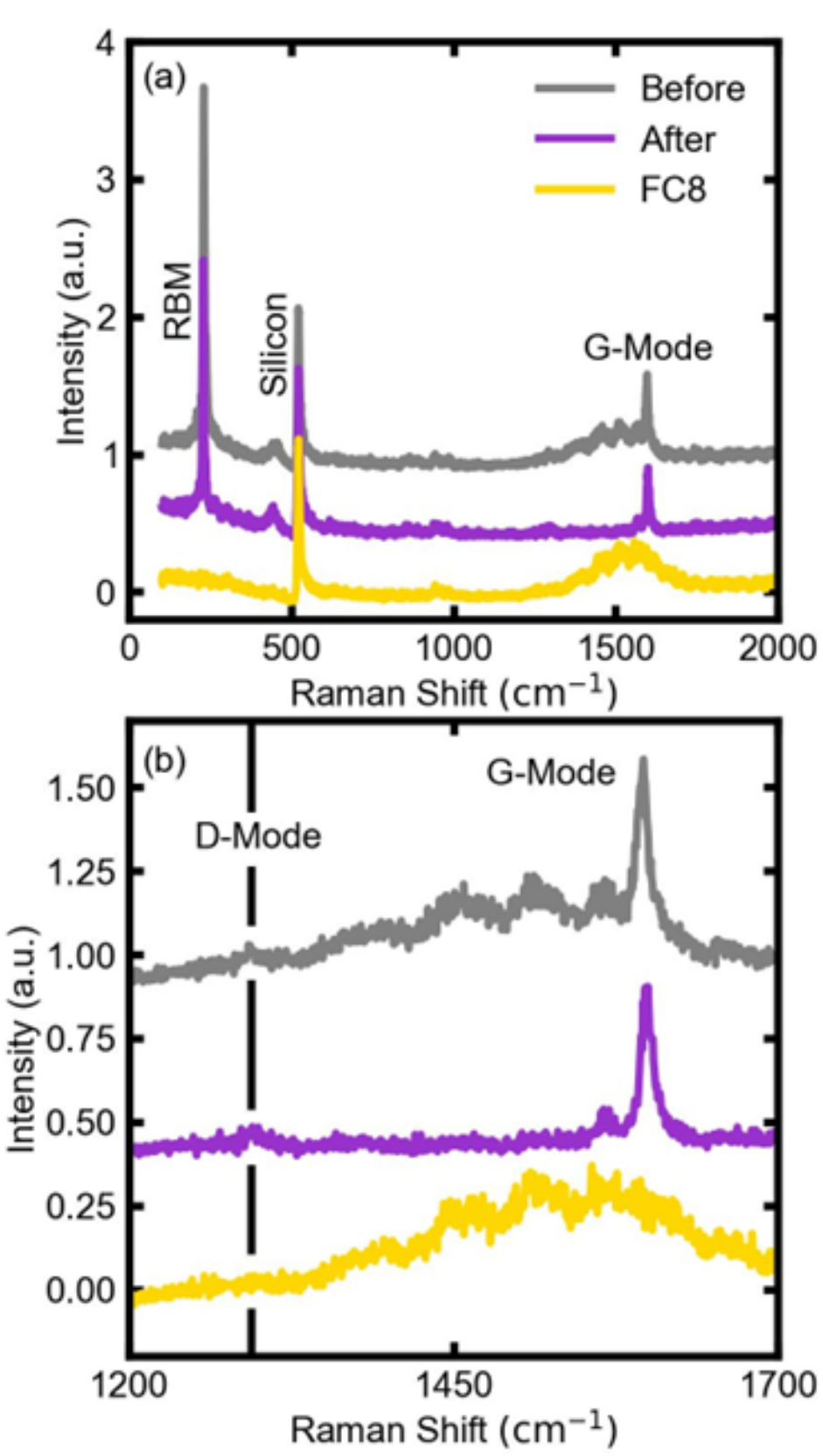

**Fig. 3.** (a) Raman spectra of a CNT-deposited wafer before (gray) and after (purple) annealing. The yellow spectrum was obtained solely from FC8. (b) Enlarged Raman spectra in the range of 1200–1700 $cm^{-1}$.

Next, CNTFETs were fabricated with a back-gate configuration on a highly doped silicon substrate. A 100-nm-thick silicon dioxide layer was used as the gate dielectric. Palladium was employed as the source and drain electrodes. The separation between the source and drain electrodes was 2 μm. In this study, CNTs were deposited after electrode patterning. This approach was selected to demonstrate the changes in device characteristics immediately before and after the cleaning process. The CNTs in these devices are exposed to air; therefore, p-type characteristics are expected[37]. The details of the device fabrication, as well as a scanning-electron-microscopy image showing a few CNTs suspended between the source and drain electrodes, are shown in supplementary data (Fig. S3).

After device fabrication, the currents flowing through the drain ($I_d$), and back-gate electrodes ($I_g$) were measured by sweeping the bias applied to the back gate ($V_g$). $V_g$ was swept from −10 V to 10 V. The source electrode was grounded, and −1.8 V was applied to the drain electrode. This measurement was performed on 245 CNTFETs. Then, the BEOL-compatible cleaning, i.e., vacuum ($<10^{-3}$ Pa) annealing at 250 °C for 12 h, was applied to the substrate. After cleaning, the electronic characteristics were measured again.

Figure 4(a) shows the $I_d$ and $I_g$ as functions of $V_g$ before and after cleaning, both obtained from the same device. As shown in Fig. 4(a), $I_g$ at 0 V < $V_g$ was below 1 pA and did not increase as much as $I_d$, which ranged from 10 pA to 1 nA. This result indicates that the current flows between the source and drain electrodes rather than between the gate and drain electrodes. Thus, the gate-leakage current, which should be distinguished from the GIDL current, is negligible in our devices. After the cleaning process, the $I_d$ value at 0 V < $V_g$ was suppressed below 0.5 pA. This value is comparable to the measurement noise floor (see Fig. S4). Therefore, the $I_{off}$ after the annealing is likely to be limited by the measurement noise floor and does not strongly depend on $V_g$. The change in the hysteresis of the $I_d$-$V_g$ characteristics was negligible, as shown in Fig. S5.

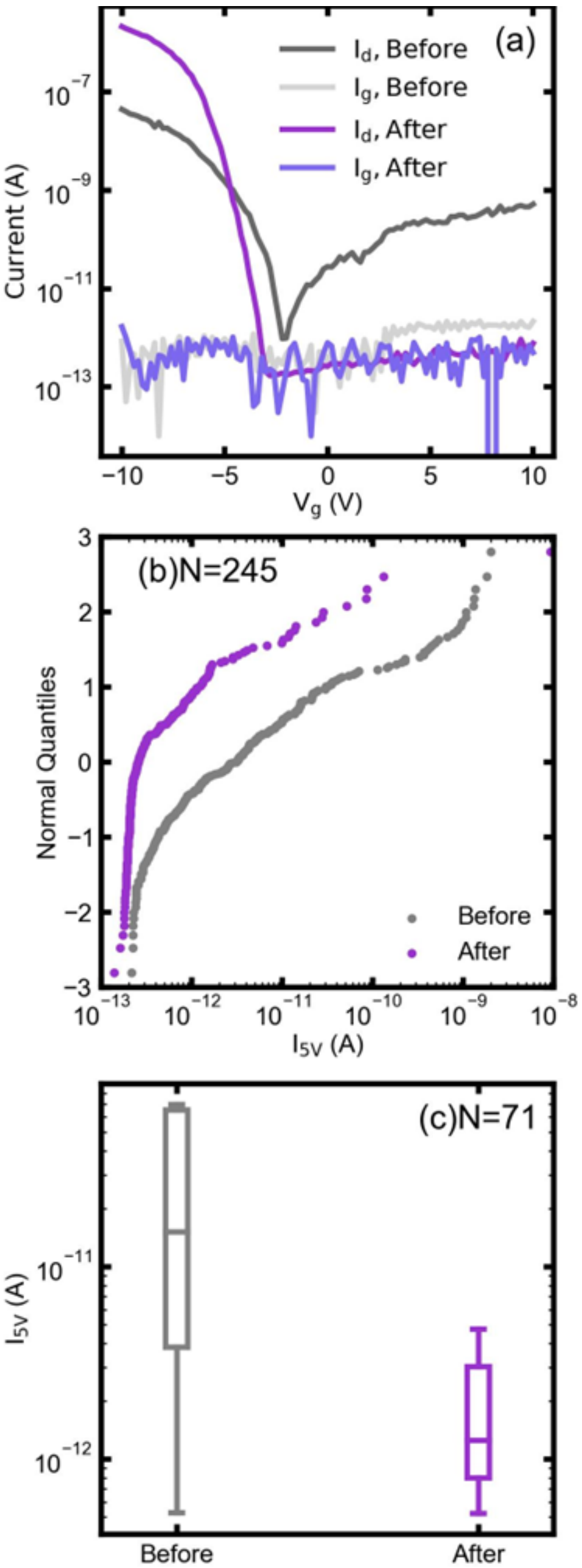

**Fig. 4.** (a) Drain and gate current as a function of gate voltage before and after the cleaning process. (b) Normal probability plots for the drain current at $V_g = 5$ V ($I_{5V}$). These data were evaluated from 245 devices on the chip. (c) Box chart for $I_{5V}$. Devices exhibiting 0.5 pA < $I_{5V}$ before and after the cleaning process were used to eliminate the limitation of the measurement noise floor.

To evaluate the effects of the cleaning process quantitatively, $I_d$ at $V_g$ = 5 V ($I_{5V}$) was evaluated before and after the cleaning process. The normal quantile plots of $I_{5V}$ before and after cleaning are presented in Fig. 4(b). As shown in Fig. 4(b), the normal quantile plot before the cleaning process weakly followed a log-normal distribution with a geometrical average of 4.0 pA. After cleaning, the distribution shifted to lower values, resulting in a geometrical average value of 0.46 pA. The distribution became steep below 0.5 pA, which may be owing to the limitation of the measurement noise floor. Therefore, to evaluate the effect of the cleaning process correctly, we selected 71 devices with $I_{5V}$ values larger than the measurement noise floor (0.5 pA) before and after the cleaning process. Box plots evaluated for the selected devices are shown in Fig. 4(c). The median decreased from 15 pA to 1.2 pA after the cleaning process, demonstrating that $I_{5V}$ decreased by one order of magnitude owing to the BEOL-compatible cleaning process. In addition to the $I_{5V}$ reduction, an increase in $I_d$ at $V_g$ = -10 V, corresponding to the maximum $I_d$, from 26.5 nA to 1.14 μA was observed. Furthermore, subthreshold slope (SS) decreased by 17%, and threshold-voltage variability decreased by 8.3% (see the supplementary data Fig. S6 and S7).

To confirm that $I_{5V}$ was dominated by GIDL, $I_d$-$V_g$ characteristics before and after cleaning were measured at different drain voltages. Figures 5(a) and 5(b) show $I_d$-$V_g$ characteristics obtained at $V_d$=-0.1 and -1.8 V, respectively. These results suggest that the observed change is unlikely to originate solely from a transition from ambipolar to p-type transport because p-type characteristics are already observed before annealing in Fig. 5(a). In addition, Fig. 5(b) demonstrates that the original $I_{5V}$ value at $V_d$ = -1.8 V was larger than that at $V_d$ = -0.1, and was suppressed below 0.5 pA after the cleaning process. These results indicate that the $I_{5V}$ value, the leakage current, depends on $V_d$. Therefore, we concluded that the $I_{5V}$ in our devices was dominated by the GIDL, and the change in $I_{5V}$ after the cleaning process was owing to the suppression of the GIDL.

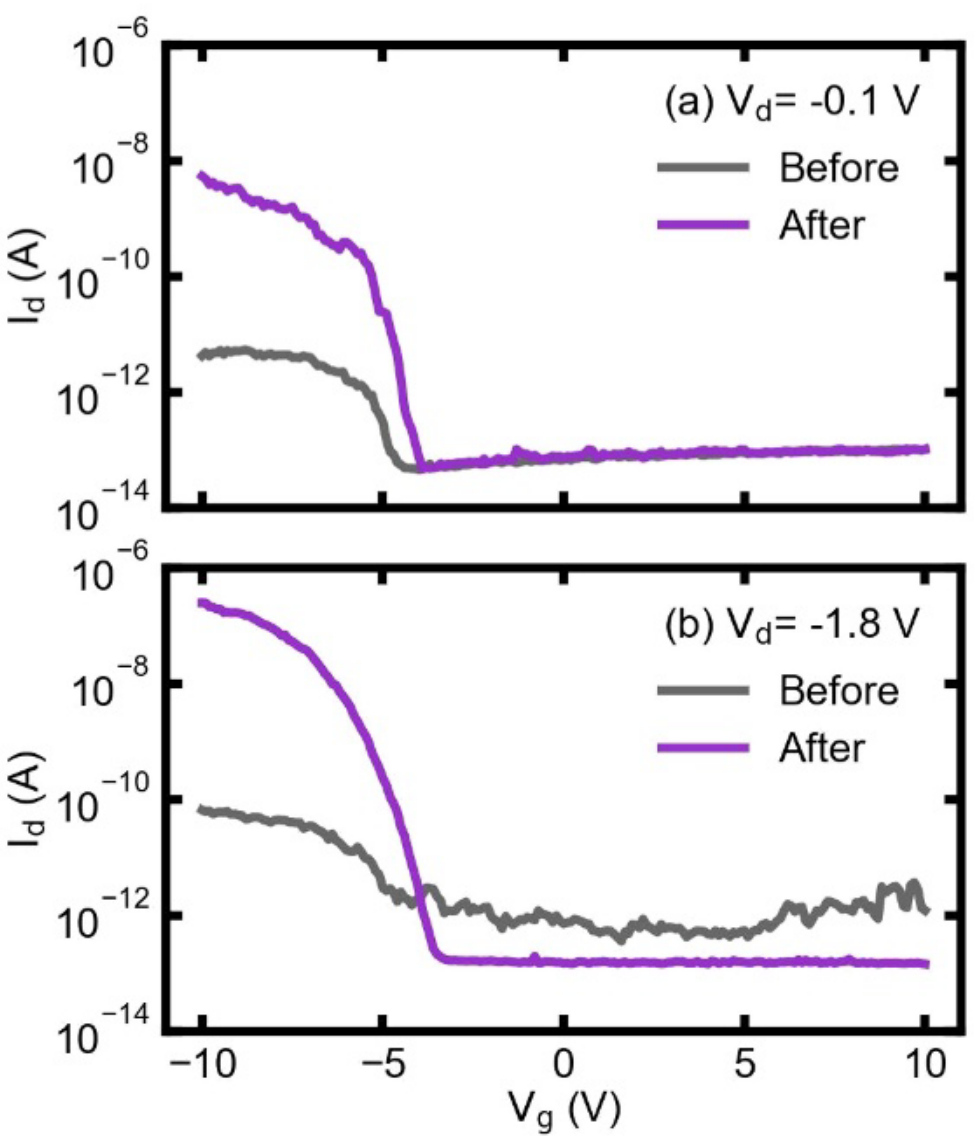

**Fig. 5.** $I_d$-$V_g$ characteristics before and after annealing at (a) $V_d$=−0.1 V and (b) $V_d$ =−1.8 V.

Next, we discuss possible scenarios in which the GIDL increases by considering the band structure of MOSFET-type CNTFETs rather than that of Schottky-barrier-type CNTFETs. This is because the Palladium used for the source and drain electrodes provides a relatively low Schottky-barrier height, and MOSFET-type CNTFETs are promising for future applications. In addition, as explained later, the proposed mechanisms are applicable to Schottky-barrier-type CNTFETs.

The GIDL is generally attributed to BTBT near the drain electrode[15]. BTBT occurs when $V_g$ is biased to a positive value and $V_d$ is biased to a negative value. Under this condition, the band structure of the CNT on the channel side shifts downward, whereas that on the drain side shifts upward. As the bias increases, band bending near the drain electrode became steeper. This resulted in a narrower tunneling width between the valence and conduction bands, and a higher GIDL. Considering that the cleaning process removes FC8 without deteriorating the CNT morphology and lattice structure, the GIDL before the cleaning process can be attributed to FC8 adsorption on the CNTs.

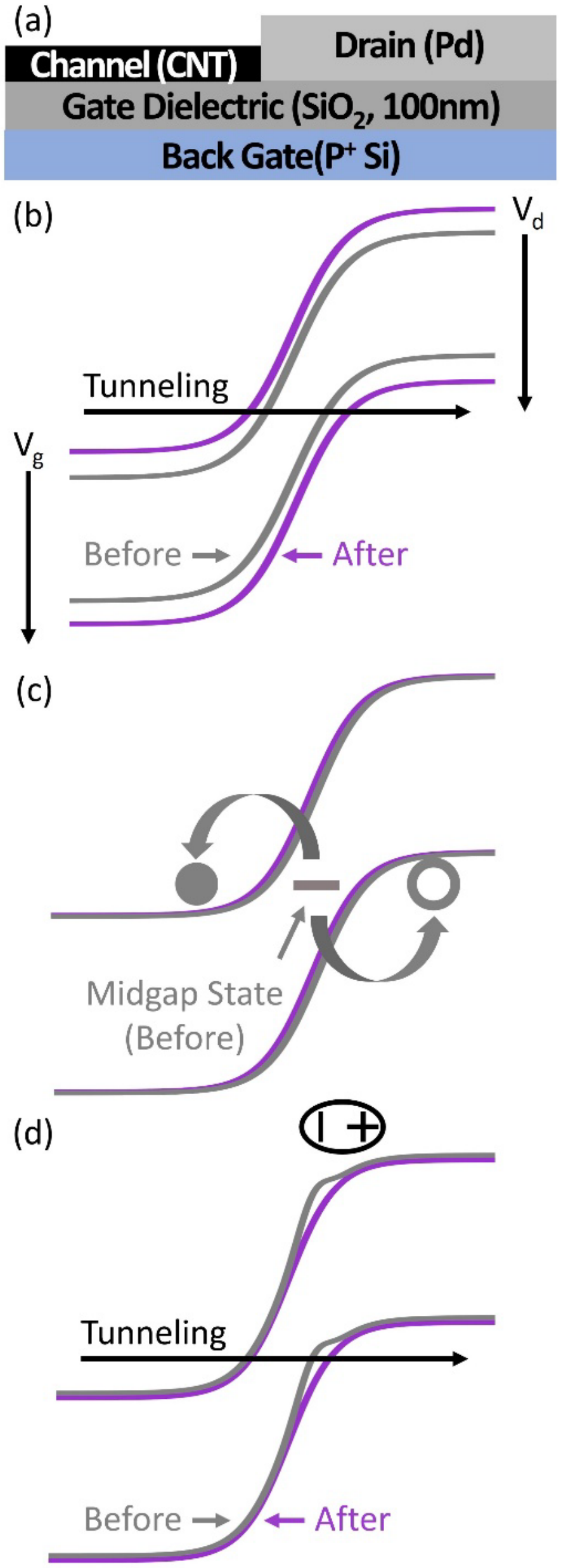

**Fig. 6.** (a) Schematic image showing the cross section of a CNTFET near the drain electrode. (b, c, d) Band structures near the drain electrode showing possible GIDL mechanisms due to the adsorption of dispersants. (b) The mechanism caused by the narrowing of the bandgap, (c) trap

assisted tunneling, and (d) dipoles located near the drain electrode. The gray and purple lines represent the band structures before and after the cleaning process, respectively.

One of the possible mechanisms resulting from FC8 adsorption is the narrowing of the bandgap [38]. Using first-principles calculations, Qiu, *et al.*, showed that several dispersants reduce the bandgap of semiconducting CNTs because the localized electronic states and orbital overlap introduced by the dispersants modify the electronic states near the band edges. The narrowing of the band gap makes the tunneling width thin and enhances GIDL as depicted in Fig. 6(b).

Another possible mechanism for the GIDL in our devices is trap-assisted tunneling (TAT), which originates from FC8 adsorption (Fig. 6(c)). In principle, if the electronic states introduced by the adsorbed FC8 are located within the bandgap of the CNTs, they may drive hole emission to the valence band and electron emission to the conduction band. This mechanism is well established for Si-based MOSFETs[39].

The other possible mechanism resulting from FC8 adsorption is the local modulation of the band structure due to the dipole interactions induced by FC8 (Fig. 6(d)). For example, when the dipole moments are locally located near the drain electrode, the dipoles moments would be oriented due to the electric field caused by the gate and drain electrodes. That is, because of the drain electrode biased to the negative voltage and gate electrode biased to the positive, dipoles would be oriented so that the positive pole points to the drain electrode. Then, the potential near the drain electrode is modulated, resulting in a small bump in the band structure of the CNTFETs (gray lines in Fig. 6(d)). These mechanisms were proposed in the prior study [40], although the origin of dipoles in the study is not dispersant but interface dipoles between metal and CNTs. In the presence of the local structure, the tunneling width between the valence and conduction bands becomes narrowed, resulting in an enhanced GIDL. Nanoscale infrared spectroscopy and mapping demonstrated finite infrared absorption by the FC8 flakes (see Fig. S8), which disappeared after cleaning. This suggests the presence of FC8-originated dipole moments that affect the local band structure of the CNTs.

After the cleaning process, these effects are expected to disappear, and the GIDL were restored as shown in purple lines in Figs. 6(b-d). These mechanisms do not necessarily occur in all CNTs within a CNTFET. When GIDL occurs, the CNTFET is off-state, and the resistance of the CNT channels is therefore high. Under such conditions, even a small number of CNTs exhibiting enhanced leakage can dominate $I_{off}$ and significantly increase the overall GIDL of the device. Also note that these mechanisms are applicable to Schottky-barrier-type CNTFETs because carrier conduction in Schottky-barrier-type CNTFETs is dominated by tunneling at the CNT-metal contacts.

If the electronic states were located within the bandgap of the CNTs, TAT was the dominant mechanism of the GIDL in FC8-adsorbed CNTFETs. Meanwhile, if such electronic states were outside the bandgap of the CNTs, GIDL mechanisms driven by the bandgap narrowing or dipole-dominated doping should be considered as the dominant mechanism. Ultraviolet photoemission spectroscopy revealed that the highest occupied molecular orbital (HOMO) level of FC8 evaluated was −6.7 eV from the vacuum level (Fig. S9). In addition, photoluminescence spectroscopy of FC8 demonstrated that the energy separation of the HOMO and lowest unoccupied molecular orbital (LUMO) levels of FC8 were approximately 2.56 eV (Fig. S10). Thus, the LUMO level of FC8 is likely located near −4.14 eV from the vacuum level. Meanwhile, the work function of the CNTs was located at approximately 5 eV[41]. With an $E_g$ of approximately 1 eV, which is a typical value for 1-nm-diameter semiconducting single-walled CNTs, the valence and conduction bands were located at 4.5–5.5 eV. Therefore, the HOMO and LUMO levels of FC8 were both energetically outside the CNT bandgap and did not provide the midgap states required for the TAT-driven GIDL. These results are contextually consistent with the prior study in which HOMO and LUMO levels of 1.3-Methyllumiflavin were estimated as –6.2 and −3.5 eV, respectively[42]. Both 1.3-Methyllumiflavin and FC8 possess the same π-conjugated core, which dominates the HOMO and LUMO levels of these molecules; therefore, the HOMO and LUMO levels of FC8 are expected to be similar to those of 1.3-Methyllumiflavin and are located outside the bandgap of the CNTs. Therefore, the bandgap narrowing (Fig. 6(b)) or dipole-dominated doping (Fig. 6(d)) are likely to be the dominant mechanisms for the GIDL in our devices. Future studies will verify the dominant mechanism of the GIDL in CNTFETs.

Finally, we briefly discuss the results of future developments in CNTFET technology. First, we considered the situation in which our device consisted of a small number of CNTs, as shown in the SEM image in Fig. S3 (e.g., three CNTs), and I5V was suppressed from 0.22 nA to 0.41 pA, as shown in Fig. 4(a). Assuming CNTFET fabrication utilizing an aligned CNT array of 400 CNTs/μm[9], our results correspond to the reduction of $I_{5V}$ from 29 nA/μm to 54 pA/μm. The latter value is as low as the $I_{off}$ value expected for high density devices[8] such as mobile, low-energy-consumption devices. Theoretically, GIDL is dominated by the local device structure near the drain electrodes rather than the channel length[43]. Experimentally, no obvious increase in GIDL was observed between L=0.5 and 0.18 μm CNTFETs[14]. Therefore, our results can be applied to moderately scaled CNTFETs. However, for highly scaled CNTFETs with a channel length shorter than 100 nm, GIDL is expected to increase because the drain electric field can more easily penetrate the entire channel region, and the degrees of design freedom near the drain electrodes to mitigate the GIDL[10] may be limited. To demonstrate the effects of the BEOL-compatible cleaning process on the GIDL for highly scaled CNTFETs, aligned CNT-array technology utilizing FC8 needs to be developed.

## CONCLUSION

In summary, we developed a BEOL-compatible cleaning process for CNTFETs based on the use of sublimable dispersants. Sublimable dispersants enabled the separation of semiconducting CNTs from metallic CNTs, while allowing the removal of dispersants through vacuum annealing at 250 °C. The cleaning process did not deteriorate the CNTs deposited on the wafers. As a result of this cleaning step, the GIDL was reduced by one order of magnitude. The result obtained in this study contributes to the development of low-power-consumption M3D devices utilizing CNTFETs. Future studies should focus on the identification of the GIDL mechanism when dispersants are locally adsorbed on CNTs and the development of aligned CNT-array technology utilizing FC8.

## SUPPLEMENTARY MATERIAL SECTION

See the supplementary material for additional information on the absorbance spectra, surface morphology, Raman mapping, device structure, statistical analysis of electrical characteristics, dielectric-dependent transport behavior, and spectroscopic characterization relevant to the electronic structure and transport mechanisms.

## ACKNOWLEDGEMENTS

This study was supported in part by the JSPS KAKENHI (grant number: JP21K04841).

The authors thank M. Takahashi, H. Ogino, Y. Zhou, S. Morita, Y. Urano, and T. Sugino for supporting material preparation and characterization.

We would like to thank Editage (www.editage.jp) for English language editing.

**AUTHOR DECLARATIONS SECTION**

**Conflict of interest**

The authors have no conflicts to disclose.

**Ethics approval**

This study did not involve any human participants or animals.

**Author contributions**

#T. I. and Y. K. contributed equally to this study.

**T. Inaba:** Conceptualization (lead), Investigation (lead), Methodology (lead), Project administration (equal), Visualization (lead), Writing – original draft (lead), Writing – review & editing (lead). **Y. Kato:** Conceptualization (supporting), Funding acquisition (lead), Investigation (supporting), Methodology (supporting), Project administration (equal), Resources (supporting), Visualization (supporting), Writing – original draft (supporting), Writing – review & editing (supporting). **Y. Iizumi:** Investigation (supporting), Methodology (supporting), Resources (lead). **Y. Fujita:** Investigation (supporting), Methodology (supporting), Visualization (supporting), Writing – review & editing (supporting). **K. Kobashi:** Project administration (equal), Resources (supporting), Writing – review & editing (supporting). **T. Morimoto:** Investigation (supporting), Methodology (supporting), Project administration (equal), Supervision (equal), Writing – review

& editing (supporting). **T. Okazaki:** Project administration (equal), Supervision (equal), Writing – review & editing (supporting).

**Data availability statement**

The data that supports the findings of this study are available within the article and its supplementary material.

# Supplementary Material for “Reducing the Gate-Induced Drain-Leakage Current of Carbon-Nanotube Transistors by Subliming Dispersants with Back-End-of-Line Compatible Temperature”

■ **Absorption spectra**

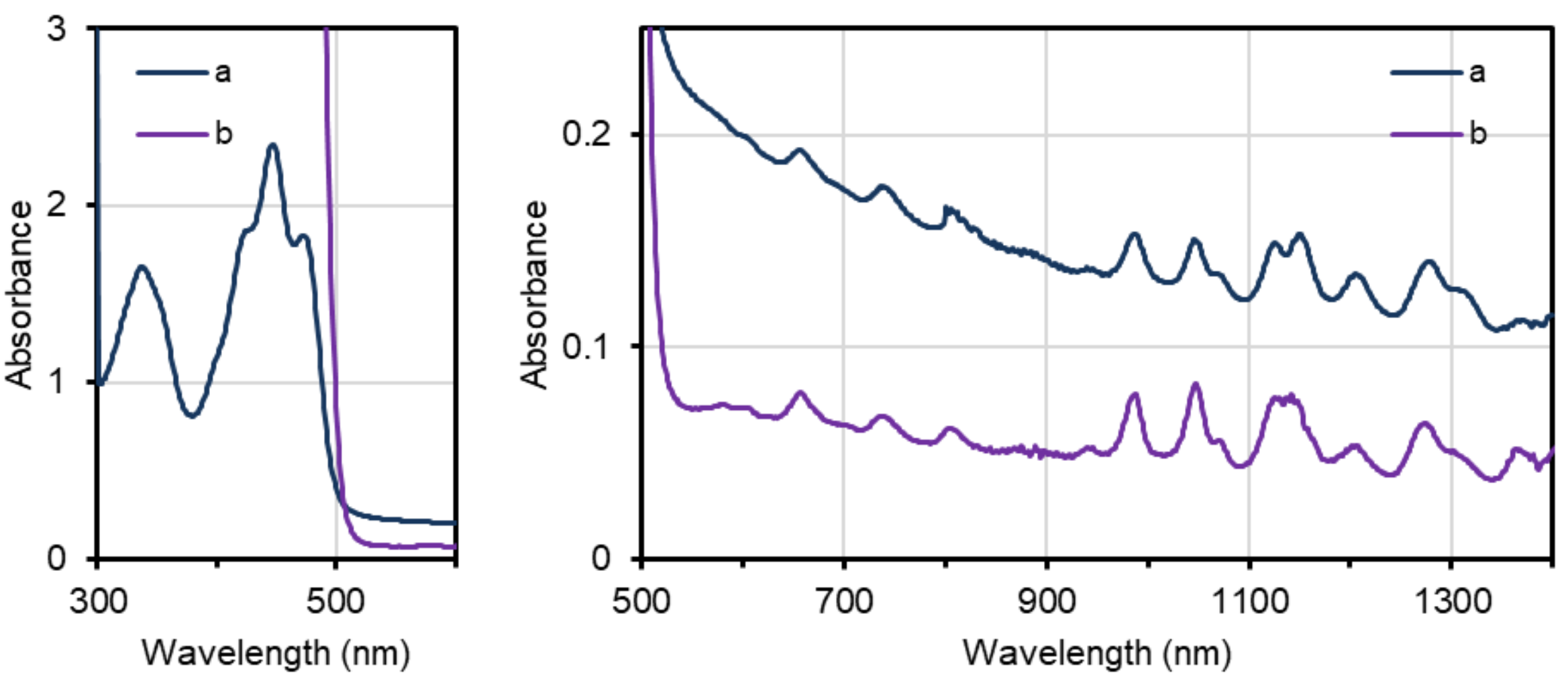

**Fig. S1**. Ultraviolet-visible-near infrared absorption spectra of CNT solutions. a (dark blue): supernatant after natural settlement for 1 h. Optical path was 2 mm. b (purple): supernatant after centrifugation at 3000 g for 10 min. Optical path was 10 mm.

Figure S1: Two absorption spectra of a CNT solution over ultraviolet to near-infrared wavelengths. The centrifuged sample shows a strongly reduced baseline while retaining distinct near-infrared peaks compared to the initial sample.

## ■ Raman Mapping

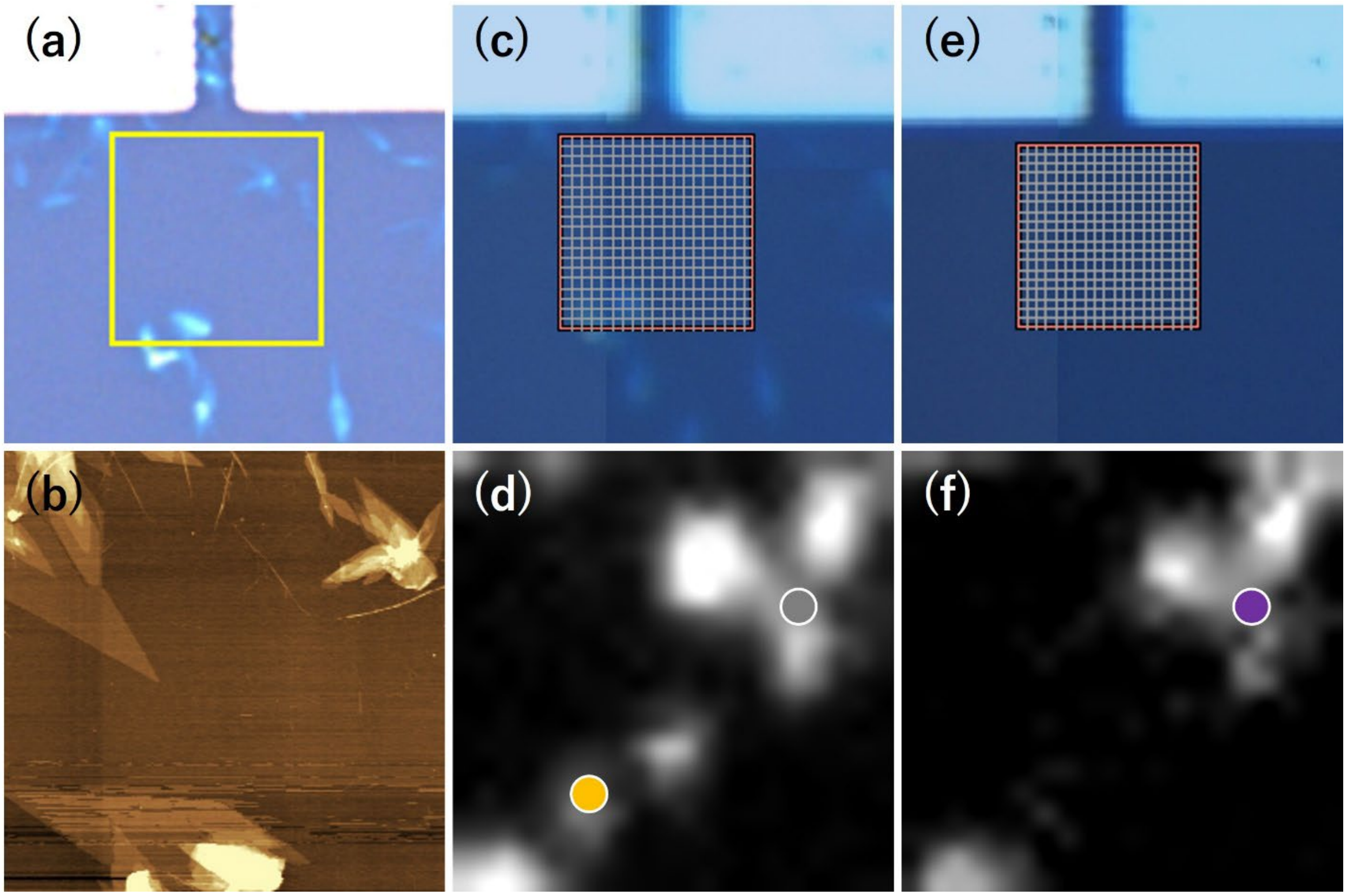

**Fig. S2**. (a) Optical-microscopy image of a CNT-deposited substrate before annealing. (b) An AFM image observed in the yellow-square area shown in (a). (c) Optical-microscopy image of the same position as (a) before the cleaning process. (d) Far-field Raman-mapping image obtained from the area outlined by the red square in (c). The Raman measurements were carried out with an incident laser with a wavelength of 785 nm. The intensity at 1590 $cm^{-1}$ is mapped. The gray dot corresponds to the position at which the Raman spectra (gray) shown in Fig. 3, which is in the main document, was obtained. (e) Optical-microscopy image of the same position as (c) after the cleaning process. (f) Far-field Raman-mapping image obtained from the area outlined by the red square in (e). Measurement conditions were the same as in (d). The purple dot corresponds to the position at which the Raman spectra shown in Fig. 3 (purple) were obtained. The bright area indicated by the yellow dot in (d) disappeared in (f). Raman signal obtained from the area originates from FC8 and is shown as the yellow lines in Fig. 3 of the main document.

Figure S2: Six images combining optical microscopy, AFM, and Raman mapping of the same CNT-deposited region before and after cleaning. The Raman maps show a localized high-intensity region present before cleaning that disappears after cleaning, while the marked measurement points confirm that the Raman spectra were obtained from the same position.

## ■ Device Fabrication

CNTFETs were fabricated with the back-gate configuration on highly doped silicon substrates (ρ=0.005–0.008 Ω·cm). A 100-nm-thick silicon-dioxide layer, which was thermally formed on a silicon substrate, was employed as the gate dielectric. The patterns of the source and drain electrodes were defined by photolithography. The separation between the source and drain electrodes was 2 μm, and the width of each electrode was 10 μm. A total of 576 pairs of source and drain electrodes were defined on 10-mm-square substrates. Then, a 150-nm-thick palladium with a 2-nm-thick titanium interlayer was deposited via sputtering. Finally, the individual source and drain electrodes were formed using a lift-off process.

After the source and drain electrodes were defined, the CNT solution was spin-coated onto the substrates. In this study, the CNTs were deposited after the source and drain electrodes were formed. This fabrication flow enabled the direct evaluation of the BEOL-compatible cleaning process. However, if the source and drain electrodes are patterned after CNTs deposition, the CNTs deposited on the substrates undergo lift-off. In this case, CNTs and FC8 could potentially be removed from the substrates, prohibiting the direct evaluation of the BEOL-compatible cleaning process.

A lift-off-free process [T. Srimani, *et al.*,VLSI Symposium 2022] will be employed for future CNTFET production. In this case, the CNT solution should first be deposited on the substrates, followed by the BEOL-compatible cleaning process. Subsequently, the source and drain electrodes will be defined.

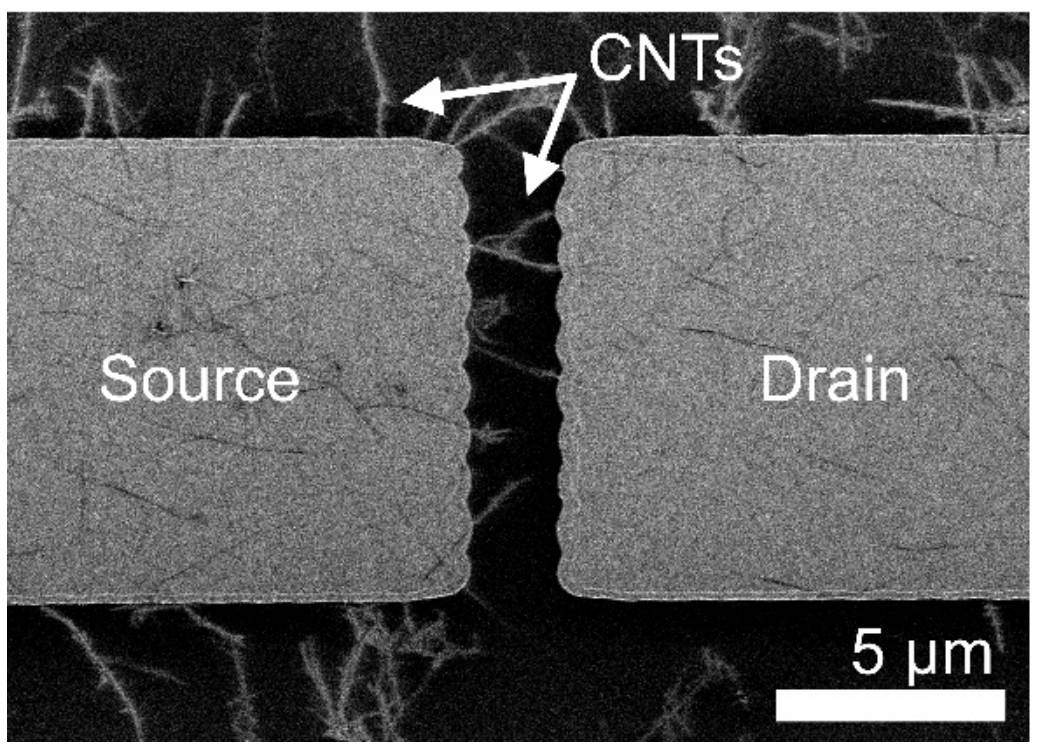

**Fig. S3**. SEM image of fabricated devices.

Figure S3: SEM image of a CNT transistor with two electrodes separated by a micron-scale gap. A few nanotubes are suspended between the electrodes, forming discrete conductive channels across the gap.

Figure S3 illustrates a fabricated device. Two CNTs that formed CNTFET channels were confirmed between the source and drain electrodes. Although densely aligned CNT arrays using sublimable dispersants have not yet been developed, such a technique is essential for the future commercialization of CNTFETs.

## ■ Measurement Noise Floor

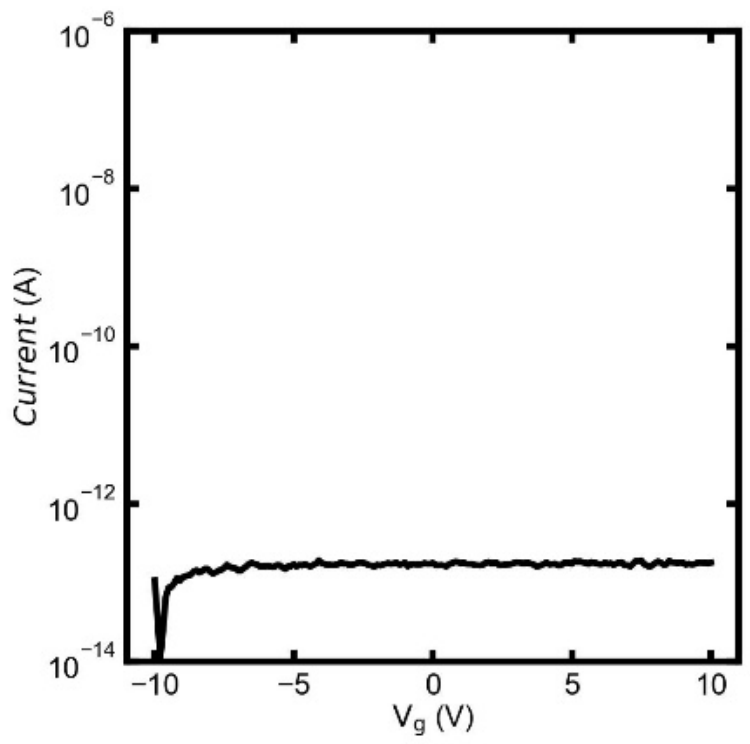

**Fig. S4.** Measurement noise floor obtained with floating probes connected to the source and drain pads of the device.

Figure S4: Current versus gate voltage measured with floating probes connected to the source and drain pads to determine the measurement noise floor. The current stays at approximately $10^{-13}$ A across the gate-voltage range from −10 V to 10 V, showing minimal dependence on gate voltage. This noise-floor level is comparable to the sub-picoampere $I_{off}$ observed after annealing.

■ **Hysteresis of $I_d$-$V_g$ Characteristics Before and After the Cleaning Process**

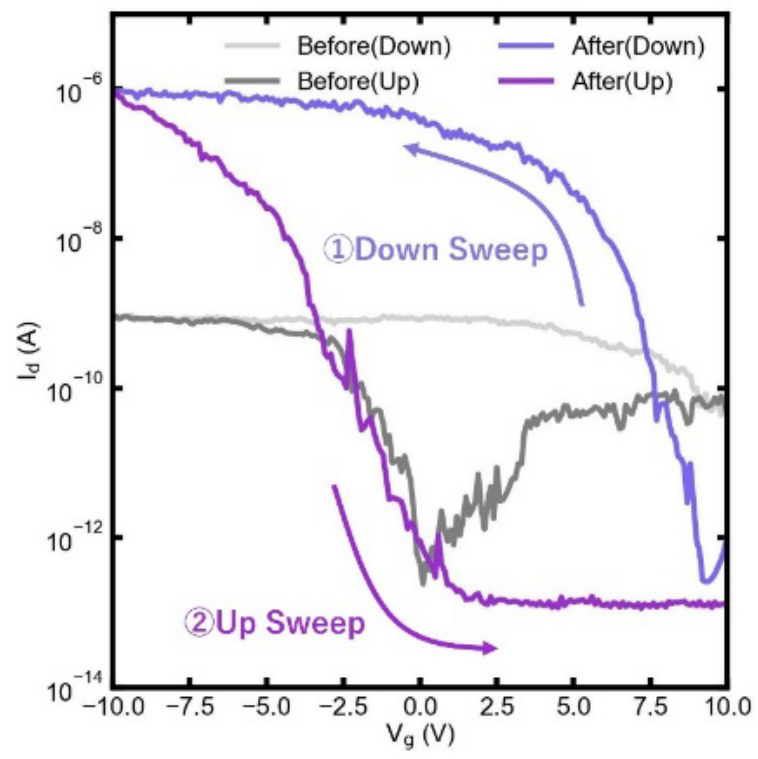

**Fig. S5**. $I_d$–$V_g$ characteristics measured at $V_d = -1.8$ V before and after the cleaning process. The hysteresis, which was observed before and after the cleaning process, is likely caused by water adsorption on the CNT surface and charge trapping in the 100-nm-thick $SiO_2$ gate dielectric.

Figure S5: Transfer curves showing drain current versus gate voltage measured with forward and reverse sweeps before and after cleaning. Similar hysteresis is observed in both cases, and the overall curve shape remains largely unchanged after cleaning.

## ■ Evaluation of Device Characteristics

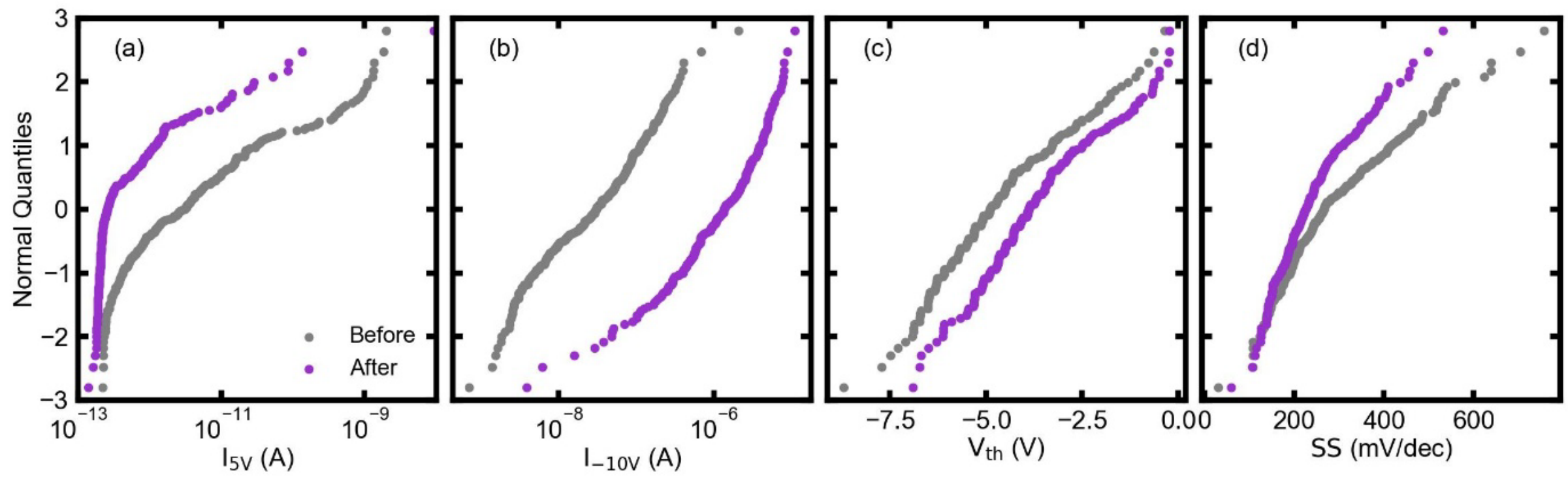

**Fig. S6**. Normal quantile plots evaluated from 245 CNTFETs. (a) Normal quantile plots for $I_d$ at $V_g$ = 5 V. The minimum value is limited by the noise floor. Geometric averages changed from 4.0 pA to 0.46 pA after the cleaning process. (b) Normal quantile plots for $I_d$ at $V_g$ = -10 V, corresponding to the maximum $I_d$ of these devices. Geometrical average changed from 26.5 nA to 1.14 μA. The increase in the $I_{-10V}$ is likely owing to the reduction of scattering centers originating from dispersants. (c) Normal quantile plots for the threshold voltage defined by the constant-current method with $I_{th}$=10 nA. The variability of the threshold voltage changed from 1.4 V to 1.3 V. The decrease may originate from the reduction of potential fluctuations on CNT channels after the cleaning process. (d) Normal quantile plots for minimum subthreshold slopes (SS) of each device. The geometrical average of SS changed from 270 mV/dec to 230 mV/dec.

Figure S6: Four normal quantile plots comparing device metrics before and after cleaning. The leakage current shifts to lower values, the maximum current shifts to higher values, the threshold-voltage distribution becomes slightly narrower, and the subthreshold slope shifts to lower values after cleaning.

■ **Analysis of Subthreshold Slope**

After the cleaning process, SS decreased from 270 mV/dec to 230 mV/dec. The SS improvement is likely due to the reduction in interface states on CNTs. That is, SS is primarily given by $(ln10) \cdot \frac{k_B T}{q} \cdot (1 + \frac{C_d}{C_{ox}})$, where $k_B$ is the Boltzmann constant, $T$ is temperature, $q$ is the elemental charge, $C_d$ is the depletion layer capacitance and $C_{ox}$ is the gate capacitance. Here, $C_{ox} = \frac{\epsilon_{ox}}{t_{ox}}$. When interface states exist, the parallel capacitance due to the interface states ($C_{it}$) is added to the depletion layer capacitance [Y. Taur, and T. H. Ning, "Fundamentals of modern VLSI devices," Cambridge university press (2021).]. Therefore, the SS can be rewritten as $SS = (ln10) \cdot \frac{k_B T}{q} \cdot \{1 + \frac{t_{ox}}{\epsilon_{ox}} \cdot (C_d + C_{it})\}$. The equation implies that the decrease in the SS after the cleaning process is caused by the decrease in $C_{it}$ and thus the reduction in interface states.

Although the SS decreased after the cleaning process, the value is still larger than the ideal value, i.e., the Boltzmann limit at room temperature $SS = (ln10) \cdot \frac{k_B T}{q} = 60\ mV/dec$, because of the $\frac{t_{ox}}{\epsilon_{ox}} \cdot C_d$ component. In particular, we employed 100-nm-thick $SiO_2$ layers as gate dielectric. The thick $SiO_2$ gate dielectric layers result in the relatively large $\frac{t_{ox}}{\epsilon_{ox}}$ and SS value compared to those obtained in prior studies employing thin $HfO_2$ layers. For example, when thin $HfO_2$ layers are employed as gate dielectric, SS values ranging from 60 to 200 mV/dec were obtained [G. Pinter, et al., 2023 IEEE Symposium on VLSI Technology and Circuits (2023)][J. Lv, et al., ACS Nano 19, 34283 (2025)][L. Liu, et al., Science 368, 850 (2020)][Y. Lin, et al., Nat. Elec. 6, 506 (2023)]. In addition, the SS value obtained by our measurements after the cleaning process (230 mV/dec) is much smaller than that reported in a prior study employing 140-nm-thick $SiO_2$ gate dielectric layers, i.e. 1,500 mV/dec [J. Appenzeller, et al., Phys. Rev. Lett. 89, 126801 (2002)]. Therefore, the SS value after the cleaning process is dominated by material and thickness of gate dielectric layers.

To show a concrete example, we fabricated CNTFETs employing 100-nm-thick $SiO_2$ and 20-nm-thick $HfO_2$ layers as gate dielectric. SiO2 layers were deposited by thermal oxidation and $HfO_2$ layers were deposited by atomic layer deposition. Other fabrication processes of these CNTFETs were the same. $I_d$-$V_g$ characteristics of these CNTFETs are shown in Fig. S6. The SS values of CNTFETs employing $SiO_2$ and $HfO_2$ layers were 298 mV/dec and 104 mV/dec, respectively. Again, these results indicate that the SS value after the cleaning process is dominated by the material and thickness of the gate dielectric layers.

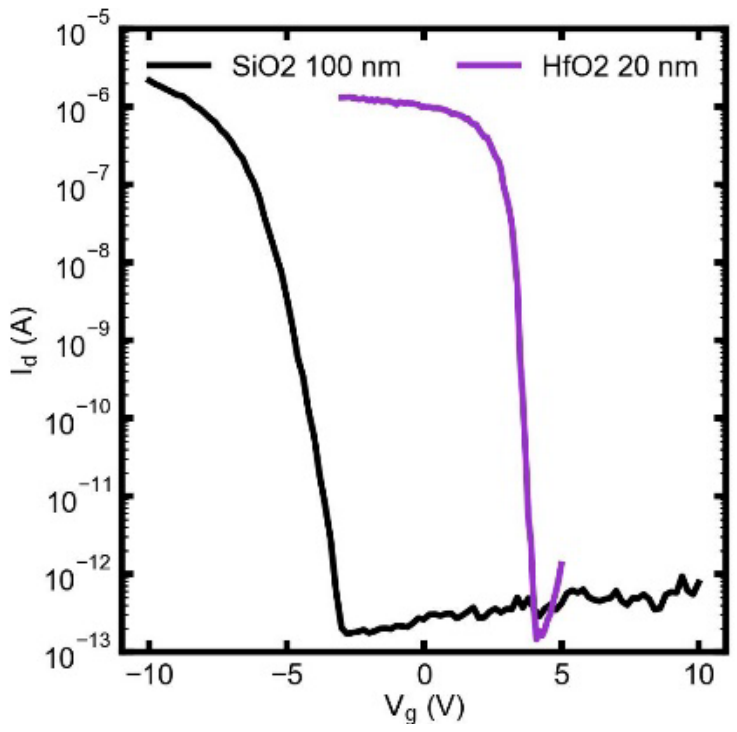

**Fig. S7**. Id-Vg characteristics obtained from CNTFETs employing $SiO_2$ and $HfO_2$ gate dielectric layers.

Figure S7: Transfer curves comparing CNT transistors with a thick oxide dielectric and a thinner high-k dielectric. The device with the thinner high-k layer exhibits a steeper subthreshold region and sharper turn-on behavior.

■ **Nanoscale Infrared Spectroscopy and Mapping**

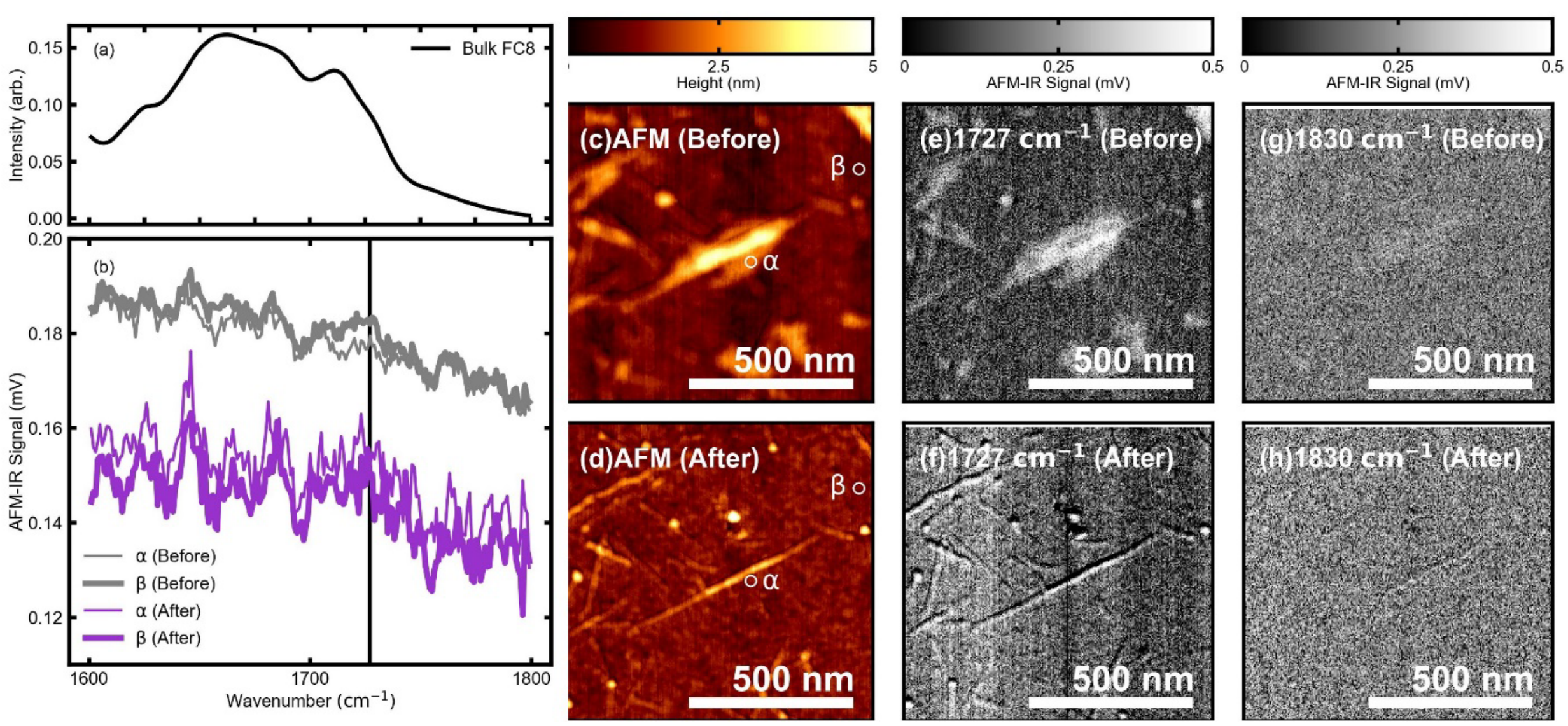

**Fig. S8.** (a) Infrared spectra obtained from bulk FC8. (b) AFM-IR spectra obtained from points α and β on the AFM images shown in (c) and (d). Gray lines correspond to AFM-IR spectra before the cleaning process. The small peak was observed at 1727 $cm^{-1}$ in the thick gray line, which corresponds to the AFM-IR spectra on point α (on the FC8 flake) before the cleaning process. However, the small peak was not observed from the thin gray line taken from point β (on the substrate). The small peak likely originates from FC8 because IR spectra obtained from bulk FC8 also exhibited a peak in the range of 1700–1740 $cm^{-1}$. Purple plots correspond to AFM-IR spectra after the cleaning process. After the cleaning process, the AFM-IR signal intensities at 1727$cm^{-1}$ obtained from points α and β were similar, suggesting that the AFM-IR signal originated from FC8 vanished owing to the cleaning process. (c, d) An AFM image of the CNT deposited substrate before and after the cleaning process, respectively. (e, f) An AFM-IR mapping image at 1727 $cm^{-1}$ before and after the cleaning process, respectively. The flake on the CNT exhibited infrared absorption, which may result in the dipole driven GIDL enhancement. (g, h) An AFM-IR mapping image at 1830 $cm^{-1}$, as control mapping images, before and after the cleaning process, respectively.

Figure S8: Infrared spectra and nanoscale infrared mapping of a CNT-coated surface before and after cleaning. A localized infrared signal present before cleaning disappears after cleaning, and the corresponding spatial map shows the loss of a bright surface feature.

## ■ Ultraviolet Photoemission Spectroscopy of bulk FC8

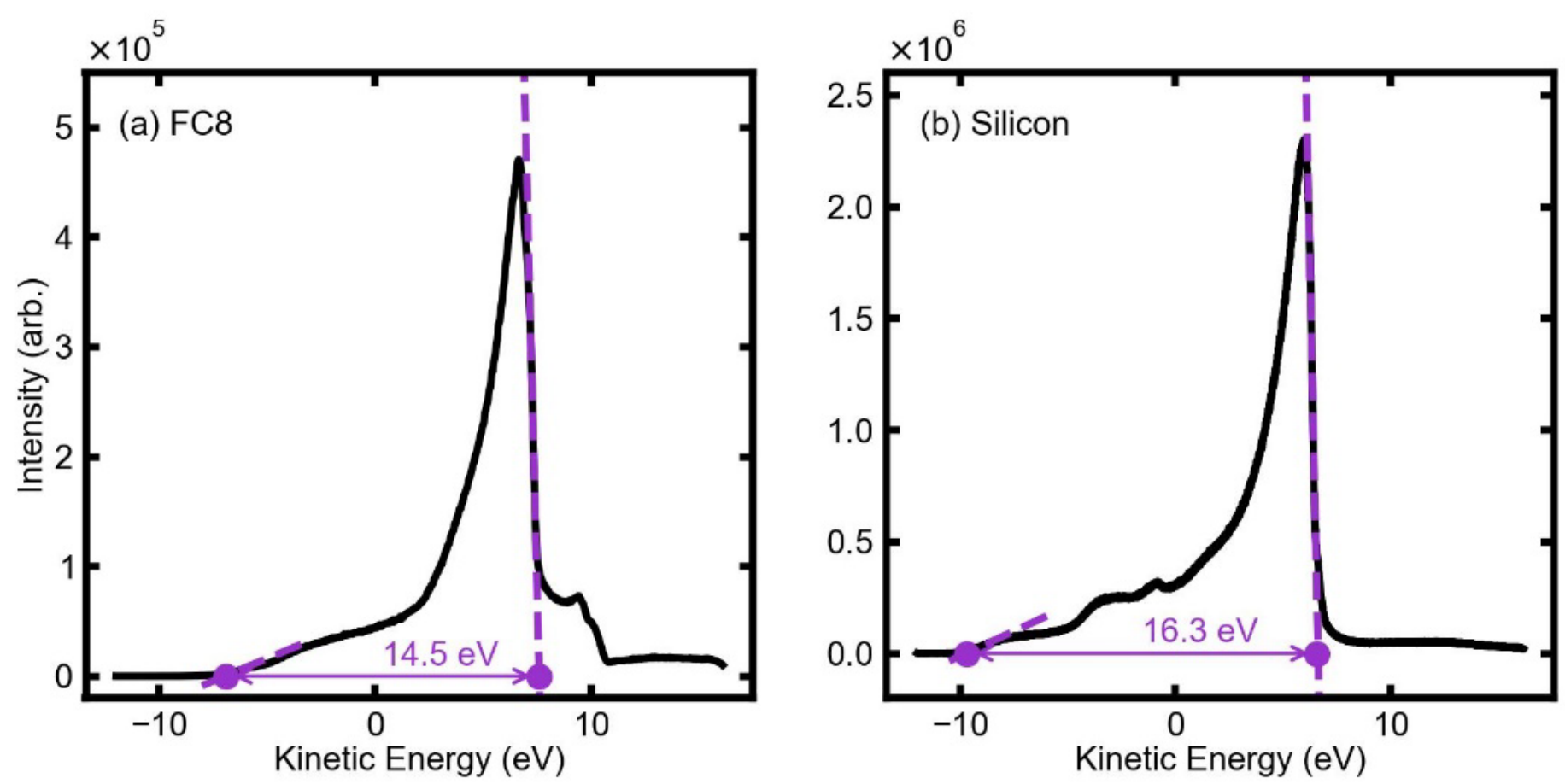

**Fig. S9**. (a) Ultraviolet photoemission spectrum obtained from bulk FC8. The spectrum was obtained using 21.2 eV excitation. The difference between the HOMO level and cutoff energy was 14.5 eV, from which the HOMO level relative to the vacuum level was estimated to be -6.7 eV. A small shoulder on the right is likely to originate from the contamination of the sample surface. (b) Ultraviolet photoemission spectrum obtained from a silicon substrate as a control spectrum. The valence-band maximum relative to the vacuum level was estimated to be -4.9 eV.

Figure S9: Two ultraviolet photoemission spectra comparing a molecular material and a reference substrate. The spectra show different cutoff positions, indicating a deeper highest occupied energy level for the molecular material.

■ **Photoluminescence Spectroscopy of FC8 in Solvent**

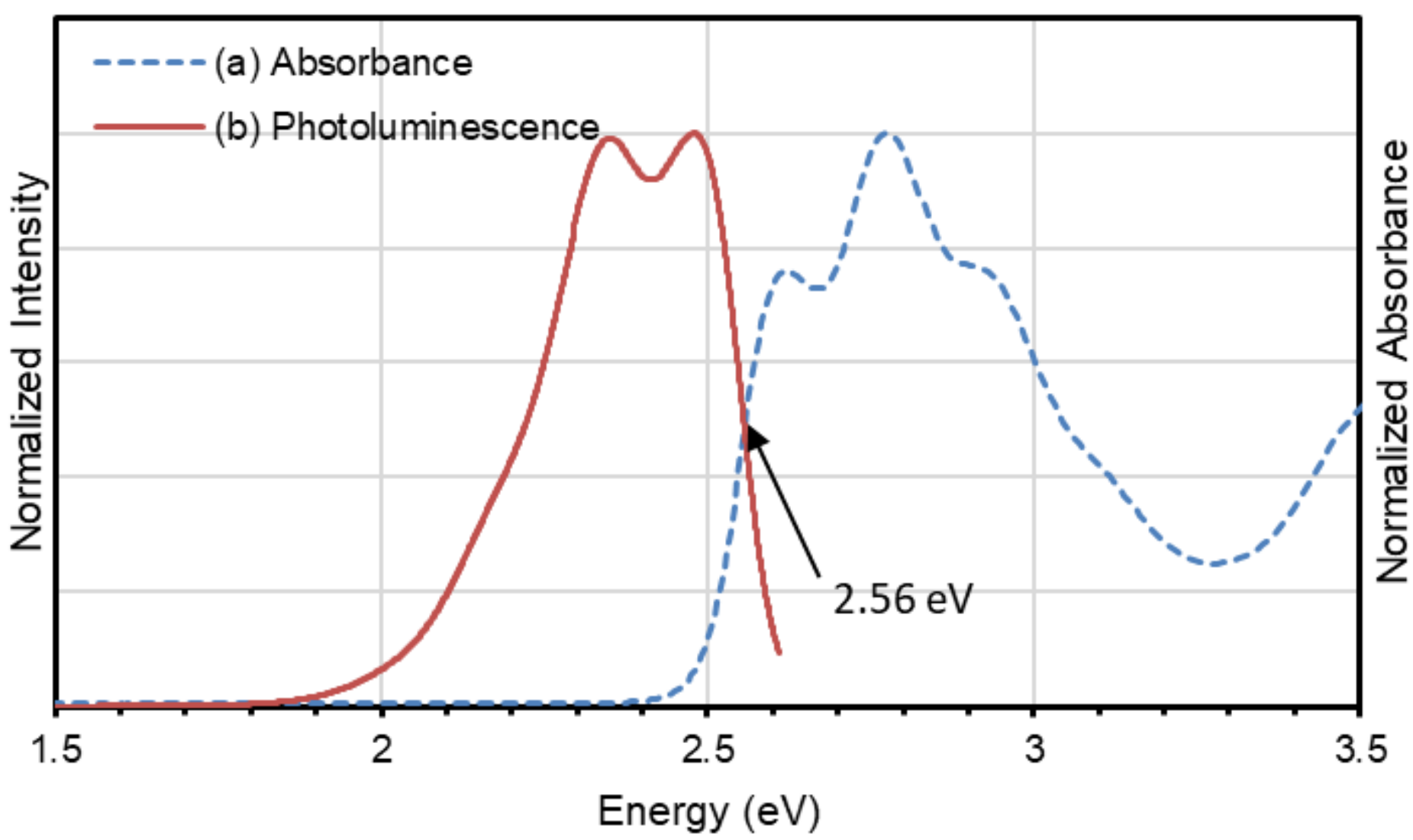

**Fig. S10**. (a) Absorption spectrum (blue dotted line) and (b) photoluminescence spectrum (red line, excitation wavelength: 445 nm) of FC8 dissolved in toluene (10 µg/mL). From these spectra, the 0–0 transition energy of FC8 was estimated to be 2.56 eV.

Figure S10: Absorption and photoluminescence spectra of a molecular sample showing separated peak positions in energy. The difference between the absorption and emission peaks indicates the optical transition energy.